\documentclass{sig-alternate-05-2015}

\usepackage{url}

\usepackage[T1]{fontenc}
\usepackage{color}

\usepackage{verbatim}
\usepackage{amssymb}
\usepackage{algorithm}
\usepackage[noend]{algpseudocode}
\usepackage{graphicx}
\usepackage{wrapfig}
\usepackage{listings}
\usepackage{mdwlist}
\usepackage{url}
\usepackage{multicol}
\usepackage{footmisc}
\usepackage[normalem]{ulem}
\usepackage{tikz}
\usepackage{lipsum}
\usetikzlibrary{automata,positioning}

\def\sharedaffiliation{
\end{tabular}
\begin{tabular}{c}}

\begin{document}

\setcopyright{acmcopyright}




\title{A Preliminary Study On Emerging Cloud Computing Security Challenges}

\numberofauthors{3} 
\author{
\alignauthor Babin Bhandari\\
       \email{babinbhandari@gmail.com}
\alignauthor  James Zheng\\     
      \email{james.zheng@deakin.edu.au}
\sharedaffiliation
      \affaddr{School of IT, Deakin University, Melbourne, Australia}
 }

\maketitle

\begin{abstract}
Cloud computing is the internet based provisioning of the computing resources, software, and information on demand. Cloud Computing is referred to as one of most recent emerging paradigms of computing utilities. Since Cloud computing is the dominant infrastructure of the shared services over the internet, it is important to be aware of the security risk and the challenges associated with this emerging computing paradigm. This survey provides a brief introduction to the cloud computing, its major characteristics, and service models. It also explores cloud security threats, lists a few security solutions , and proposes a promsing research direction to deal with the evolving security challenges in Cloud computing.
\end{abstract}




\section{Introduction}
Cloud computing is the current advancement in the field of information technology. During last couple of years cloud computing has come under the spotlight~\cite{hamlen2012security} as the major trend in the field of information technology. The cloud computing is the concept where collection of services and resources required all over the world are distributed from the data centers over Internet. In recent years, cloud computing is widely used for delivering computer resources (e.g., storage, networks, applications, and servers) as a services. Google AppEngine, Microsoft Azure, Amazon AWS, IBM Bluemix are some of the major players offering cloud computing services. The innovative way of delivering services and resources on demand has dramatically changed the way organization are operated. 
Based on~\cite{mell2011nist,weissman2009design}, we list a few essential characteristics of cloud computing:
1. On Demand Service, which allows consumers to use cloud computing resources as needed;
2. Broad Network Access, which allows consumers to access cloud computing resrouces whenever and wherever as required;
3. Resources Pooling, where multiple consumers share pool of computing resources such as storage, applications, memory, storage, network bandwidth, virtual resources etc. over the internet that are made available by the service providers. The physical and virtual resources of the service providers are assigned and reassigned as per the consumers demand. It assists different large scale and medium scale organization to obtain access to the large pool of resources rather than maintaining their own physical or virtual infrastructure;
4. Rapid elasticity, which allows scaling resources both up and down as needed by the customers quickly;
5. Measured Service, which means that resources usage can be easily measured, controlled reported, and optimized automatically. 

Beside all the positive sides of cloud computing,  there exist security barriers that need to be properly addressed~\cite{hamlen2012security,rai2013securing}. In this report, we first walk through existing service models and talk about some securities issues with these models.  Then we list some of the major security challenges associated with cloud computing in general.  Finally we propose some solutions to deal with these security challenges.

\section{CLOUD Service Models  }
In this section, we will walk through some popular cloud service models and summarize the security mechanism used in each model.

\subsubsection{Infrastructure as a Service (\emph{IaaS})}
Cloud infrastructure such as computing resources, data storage, firewall, networks, severs, load balancers are the core components of IaaS. It can also be said that computing resources such as hardware and software components providing service to the cloud end users/consumers is known as infrastructure as a service. Cloud consumers can subscribe to any of these services wherever and whenever required and are charged as per the usage of the resources and time. The cloud service makes these resources available to different sized organizations and removes the need for these business
to maintain their own resources. 

IaaS only provides basic security such as firewalls, but to make customers completely reliable on the IaaS service providers,  more comprehensive, reliable, and systematic security guarantee (e.g., intrusion detection system) must be placed.

\subsubsection{Platform as a Service (\emph{PaaS})}
In this cloud delivery model,  the cloud service providers offer an online platform to the cloud application developers for software development. 
Some of the example of platform provided by the cloud service providers are Amazon AWS, Force.com, Microsoft Azure, and Google App Engine. 
PaaS provides flexibility and simplicity that allows quick implementation without having to make much configuration~\cite{sarodesecurity}.

Major PaaS providers offer the following secuirty protection:  
\begin{itemize}
\item Protecting VM against malicious attacks 
\item Maintaining integrity of the application 
\item Authentication checks during file transfer 
\end{itemize}

Similar to the security protection offered by IaaS, there are no trust worthy and systematic security mechanism here (e.g., systematic encryption or intrusion detection systems) to help application developers secure mission critical cloud applications (e.g., finanical applications).

%

\subsubsection{Software as a Service (\emph{SaaS})}
The core components of this service models are email, CRM, virtual communication, gaming applications running on the cloud infrastructure. This cloud service model facilitates the consumer to have an effective access to the needed applications running on the cloud with different client devices through client interfaces such as web browsers.  
In a nutshell, SaaS is typically a software distribution model where a number of applications are hosted by the cloud service provider and are made available to the consumers over Internet. 

Some of the methods of securing SaaS applications can be listed as below  
\begin{itemize}
\item SSL- Secure Socket Layer  
\item XML Encryption 
\item Web Services (WS) security 
\end{itemize}
Similar to the security analysis to IaaS and PaaS, these methods are far less enough to guarantee the security requirement of a mission critical SaaS application offering to multiple tenants, where security and privacy demands are even higher than a normal cloud application.

\section{Security Challenges}
In this section, we list some of the security challenges in Cloud Computing below.

\subsection{Data Confidentiality} 
Confidentiality ensures that the sensitive data and information are not disclosed to unauthorized individual, devices or the system. In the cloud computing environment the computing resources (e.g., services, storage, servers, and database) reside in the cloud service provider's premises [20]. 
This sharing of computing resources with a number of users increase the secuirty risk and possibly violates the confidentiality of each customer. Since the location of the data is only known to the service providers, the service provider is fully responsible for making the users data confidential and ensures that the customer's information cannot be accessed by other users on the cloud systems.

\subsection{Data availability }
Cloud consumer is fully dependent upon the cloud service provider for their data and information as they store their information at remote location owned by the service providers. Sometimes the system failure on the providers end can make it very difficult for the data owner to access the data at the required time. 
Any type of flooding attacks (e.g., Denial of Service) can result in the interruption of the service for a certain period of time. 
  
\subsection{Data Integrity}
Data integrity ensures the quality, correctness, consistency, and the accessibility of the user's data in cloud environment. Since the user's data and information are stored remotely on the service providers premises,  any compromise of the server or the resources may result in the data alteration and the corruption~\cite{masky2015novel}. So the cloud service providers should offer monitoring of the data integrity in order to minimize possibilities of data corruption and crash. The cloud service provider is fully responsible for maintaining the data integrity for the user's data in the cloud environment. 

\subsection{Accessibility}
Cloud service applications are provided over the internet,
 which makes the accessibility of the cloud services a lot easier from different devices (e.g., laptops, mobile devices, Tablet, and PCs). Though the wide accessibility of service improves convenience for customers, it further exposes the cloud service to the security risk and threats. The report~\cite{sirohi2015cloud} outlines mobile computing threats such as information stealing, insecure network (Wi-Fi), OS vulnerability, and insecure marketplace.

\subsection{Data Security}
In the cloud environment the enterprise or the individual's data are stored in the service provider's premises, thus it is up to the service providers to have strict access control guidelines. Cloud service providers must ensure the security by implementing additional security, frequent security audits, security check,  risk assessment, and risk management plans in order to protect data security on behalf of customers~\cite{zhao2014cloud}.

\subsection{Data location}
As the cloud consumer is unknown and unaware about action location of their own data in the cloud environment, this sometime may result in a legal issue if there are any serious compromise of data security.

\subsection{Identified Security Threats}
Since cloud serivce provides store customers data in their own premise, customers must have the trust with the providers. However number of threats present in the cloud computing environment seriously undermines this trust.
Table.~\ref{table:12Threats} summarises the critical threats to cloud security identified in existing studies~\cite{chen2012data}. Without solving these security issues, enterprises (mainly those large ones) would be very hesitatiing to migrate their application and data to the cloud.
\begin{table*}[!ht]
\centering
\caption{Cloud Security Threats  }
\begin{tabular}{|c|l|} \hline
Threat &	Description \\ \hline
Data Breach	& Condition where the sensitive data is released, stolen, viewed\\ 
& or used by unauthorized person or the system \\ \hline
Insufficient identity, credential and access management& 	Lack of the flexible and up-to date  certificates,\\ & weak passwords, unsecure cryptographic keys\\ \hline
Insecure interface and APIs&	Exposure of applications programming interfaces and software\\ 
& user interfaces utilized by cloud services  \\ \hline
System vulnerabilities&	System vulnerabilities within the operating systems kernel,\\ 
& software's or libraries  can  risk cloud services \\ \hline
Account Hijacking &	Use of common methods like phishing, vulnerability exploitation to gain\\
& account credential and using them to exploit cloud.\\ \hline
Malicious Insiders	&Former employee, contractor or partner who had or has the\\
& access and intentionally target the system.\\ \hline
Advance persistent threats	&Sophisticated attack planned over \\
& extended time period \\ \hline
Data loss	&Loss of data due to malicious attacks or accidental \\
& deletion or physical catastrophe\\ \hline
Insufficient due diligence	&Insufficient roadmap and the checklist for the\\ 
& great chance of success.\\ \hline
Abuse and nefarious use of cloud services& 	Expose of cloud environment due to unnecessary use of its \\
& service because of its weak security standards. \\ \hline
Denial of service &	Attack that causes cloud services an intolerable system \\
& slowdown or even a service unavailability. \\ \hline
Shared Technology issues& 	Sharing of hardware components leads to the compromise across the \\
& entire deployment model if there exist a single misconfiguration.\\ \hline
\end{tabular}
\label{table:12Threats}
\end{table*}

\section{Proposed Solution}
There are a few existing research works providing a clear view of security issues and to ensure the data security and trust in cloud computing environment. Although, cloud computing is still in its advancement phase, there have been numbers of approaches and methods proposed by researchers for the improvement of cloud computing security. 
In ~\cite{masky2015novel}, the study proposed a novel risk identification framework for cloud computing security. 
The study also showed that the identification of risks and threats is one of the important aspects for security of cloud environment. Hence, the authors presented the OCTAVE approach (Operationally Critical Threat, Asset, and Vulnerability Evaluation) as a novel framework to assess the risk of cloud computing. OCTAVE approach focus to provide the extensive risk assessment of operational cloud computing environment and produce more robust framework. The framework is designed by focusing on where the cloud data and information assets are stored, how they are accessed, transmitted, processed and how the cloud computing environment is exposed to threats and vulnerabilities. 

In~\cite{sirohi2015cloud}, it proposed an approach for data security assurance in cloud with encryption and decryption techniques. The paper presents the security framework with utilization of Proxy and Homomorphic based encryption methods, which has been assisted with the malware detection, real time data monitoring,and forensic virtual machine. 
Similarly, Cloud computing security solution with the use of Homographic encryption has been discussed in~\cite{zhao2014cloud}. The proposed security solution focused on providing the security of data during transmission and storage in cloud computing infrastructure with the use of Homomorphic encryption technique. Furthermore the paper provides the solution that is fully applicable for the retrieval and processing of encrypted data.  In~\cite{daniel2014challenges}, the study proposed the series of policies like information policy, security policy, audit policy, access control policy, and risk control policy for the better reliability and privacy in cloud computing environment. 

With the advancement of Microservice-architecture enabled cloud applications~\cite{zheng2018smartvm, zheng2017big, zheng2017smartvm, zheng2017bigvm, pham2017paas}, there are studies to investigate the security vulnerabilites with the new paradigm.  In~\cite{zeng2018aua, xu2018csp}, new efficient  authentication and contract signing protocols are investigate with low-resource usuage and communication latency.  In~\cite{yu2018survey}, comprehensive study has been conducted mainly toward security issues in the Microservices communication.

In summary, most of the previously proposed solutions presented above seem to focus on specific aspect of the security and privacy issues of the cloud computing environment. Cloud computing is a broad topic with equally broad security issues and surrounding threats and challenges. In order to protect the cloud computing service providers and cloud consumer, effective runtime verification~\cite{zheng2015braceassertion,zheng2016efficient,zheng2014physically,pan2017cyber,zheng2017security} along with anomaly and signature based intrusion detections system are strongly recommended to handle the emerging and evolving security threats in cloud computing~\cite{sangeetha2015signature,thu2013integrated}. Anomaly based intrusion detection system analyzes all the network traffics and detects the abnormal behaviors of network packets and applications. Anomaly based systems are considered to be a technique that is able to protect target systems from malicious attacks. Whereas, signature based intrusion detection system detects attacks and intrusions based on the predefined sets of signatures. Hence it can be understood as a system that detects the known attacks. Therefore, Combination of anomaly and signature based intrusion detection system can be a good cloud security solution that can identify and detect both known as well as unknown security threats. However, proper network intrusions detection algorithm that is able to detect all the previously known and unknown threats with minimum false alarms is still urgently required.

\section{Conclusion}
Although the cloud computing focuses to provide cost effective and timely service to the consumers, there exist some serious security challenges that need to be addressed. Especially the data privacy and integrity of the customer's data on the cloud service providers premises are the key security concerns.  In this paper we identified the major security challenges in the cloud computing, lised some state of the art security mechanism in dealing with these challenges, and propose a promising combined intrusion detection system as a future work.

\bibliographystyle{abbrv}
\bibliography{security}  

\end{document}